\documentclass[12pt,preprint]{aastex}		

\newcommand{\kms}{km\,s$^{-1}$}
\newcommand{\ms}{m\,s$^{-1}$}
\newcommand{\hr}{HR\,3831}
\newcommand{\vb}{\mbox{\boldmath$V$}}

\shorttitle{Indirect Imaging of Stellar Pulsations}
\shortauthors{Kochukhov}

\begin{document}

\title{Indirect Imaging of Nonradial Pulsations in a Rapidly Oscillating Ap Star\footnotemark[1]}
\footnotetext[1]{Based on observations obtained at the European Southern Observatory, La Silla, Chile}

\author{Oleg Kochukhov}
\affil{Institut f\"ur Astronomie, Universit\"at Wien, T\"urkenschanzstra{\ss}e 17, 1180 Wien, Austria}
\email{kochukhov@astro.univie.ac.at}

\begin{abstract} 
Many types of stars show periodic variations of radius and brightness,
which are commonly referred to as `stellar pulsations'. Observed
pulsational characteristics are determined by fundamental stellar
parameters. Consequently, investigations of stellar pulsations provide
a unique opportunity to verify and refine our understanding of the
evolution and internal structure of stars. However, a key boundary
condition for this analysis -- precise information about the geometry
of pulsations in the outer stellar envelopes -- has been notoriously
difficult to secure. Here we demonstrate that it is possible to solve
this problem by constructing an `image' of the pulsation velocity field
from time series observations of stellar spectra. This technique is
applied to study the geometry of nonradial pulsations in a prototype
magnetic oscillating (roAp) star \hr. Our velocity map directly
demonstrates an alignment of pulsations with the axis of the global
magnetic field and reveals a significant magnetically induced
distortion of pulsations. This observation constitutes a long-sought
solution of the problem of the pulsation geometry of roAp stars and
enables very stringent tests of the recent theories of stellar
magneto-acoustic oscillations. 
\end{abstract}


\section{Introduction}

A classical solution \citep{uoa89} of hydrodynamic equations expresses
monoperiodic nonradial pulsational fluctuations of an idealized spherically
symmetric star in terms of a spherical harmonic function. Approximations of
this simple stellar pulsation theory are adequate for a significant fraction of
oscillating stars and, therefore, historically investigations of stellar
pulsation geometry were limited to assigning spherical harmonic `quantum
numbers' $\ell$ and $m$ to each observed pulsation mode. This problem is solved
by several varieties of the mode identification technique \citep*{apw92,w88},
which, with different degrees of success, are applied to photometric and
spectroscopic observations of pulsating stars. However, pulsation modes in real
stars are often significantly distorted by rapid stellar rotation \citep{ls90}
and global magnetic field \citep{dg85,bd02,sg04}. These effects result in a
considerably more complicated pulsation geometry, irreducible to a single
spherical harmonic function. Up to now no objective technique was available for
a comprehensive and reliable analysis of such distorted nonradial pulsations,
but it was understood that a detailed study of  high-resolution stellar spectra
is the most promising approach to resolve these difficulties.

In general, all inhomogeneities in the outer layers of stellar envelopes,
including pulsational disturbances, produce distortions in the stellar line
profiles. As a star rotates and starspots appear and disappear from the visible
hemisphere, spectral line shapes and strengths are modulated periodically.
Doppler imaging (DI, e.g., Vogt, Penrod, \& Hatzes \citeyear{vph87}) is a powerful
technique to decipher rich information about the physics and geometry of the
surface structures contained in the stellar spectrum variability.  The Doppler
mapping method is based on the resolution of the stellar surface provided by
the effect of stellar rotation on the shapes of absorption spectral lines. If
the projected rotational velocity of a star is large enough for the shapes of
spectral lines to be dominated by the rotational Doppler broadening, there
exists a well-defined mapping relation between distortions in a line profile
and positions of the corresponding inhomogeneities on the stellar surface.
Thus, a single Doppler-broadened line profile represents a one-dimensional
projection of the visible stellar hemisphere, with all the structure blurred
along the lines of constant Doppler shift. Spectroscopic observations at
different rotation phases can be used to monitor the evolution of dips and
bumps in line profiles, thus obtaining a set of non-redundant projections of a
stellar surface map, and eventually to reconstruct a two-dimensional image from
a time series of line profile observations. This indirect stellar surface
imaging method was first applied to the problem of mapping  chemical
inhomogeneities in early-type stars \citep{krw86} and was subsequently extended
to investigation of temperature spots \citep{vph87} and complex magnetic
topologies \citep{s89} in active late-type stars and global magnetic fields
\citep{pk02} in chemically peculiar stars.  \citet{vp83} have emphasized that
rotational broadening of stellar spectra  also aids in probing the horizontal
structure of nonradial pulsations. However, subsequent spectroscopic studies
of rapidly rotating oscillating stars \citep{gk88,kwm92,h98} used only
resolution of pulsation structures along stellar longitudes and restrictively
approximated velocity field with a single spherical harmonic.

In a recent development Berdyugina, Telting, \& Korhonen (\citeyear{btk03}) presented a
new exploration of the capabilities of the DI technique in its application to mapping
of stellar nonradial pulsations. These authors employed a temperature DI code in
modeling pulsational line profile variations and assumed that pulsational fluctuations
at the stellar surface can be represented by a rotation of a fixed pattern. As
emphasized by \citet{k04}, the latter assumption restricts pulsational mapping to the reconstruction
of a spherical harmonic geometry, which is aligned with the stellar rotation axis and
is characterized by a unique non-zero azimuthal number $m$. This condition is not
fulfilled for zonal modes and is invalid for all types of oblique stellar pulsations,
hence the method  of \citet{btk03} cannot be applied to study oscillations of roAp
stars. 

Here we present the first application of an alternative technique. 
In our approach the principle of Doppler imaging is extended to the
reconstruction of the time-dependent velocity field. The temporal and surface variation
of the velocity vector are represented with a superposition of the two constant
surface distributions: 
\begin{displaymath}
\vb(t,\theta,\phi) = \vb^c(\theta,\phi) \cos(\omega t) + 
\vb^s(\theta,\phi) \sin(\omega t),
\end{displaymath}
where $\omega$ is the pulsation frequency and
$\theta$, $\phi$ are usual spherical coordinates on the stellar surface. The $\vb^c$
and $\vb^s$ vector maps are recovered directly from the observed line profile
variability without imposing any specific global constraints on the pulsation geometry.
This is equivalent to mapping a two-dimensional surface distribution of the pulsation
amplitude and phase for each velocity component. 

The foundations of the pulsational Doppler mapping and description of
its computer implementation were presented by \citet{k04}. We refer the reader
to this paper for a detailed explanation of the technique and discussion of
the numerical simulations which were used to evaluate performance and intrinsic
limitations of the new surface mapping method. 

\section{Pulsational Doppler mapping of \hr}

\begin{figure}
\epsscale{0.5}
\plotone{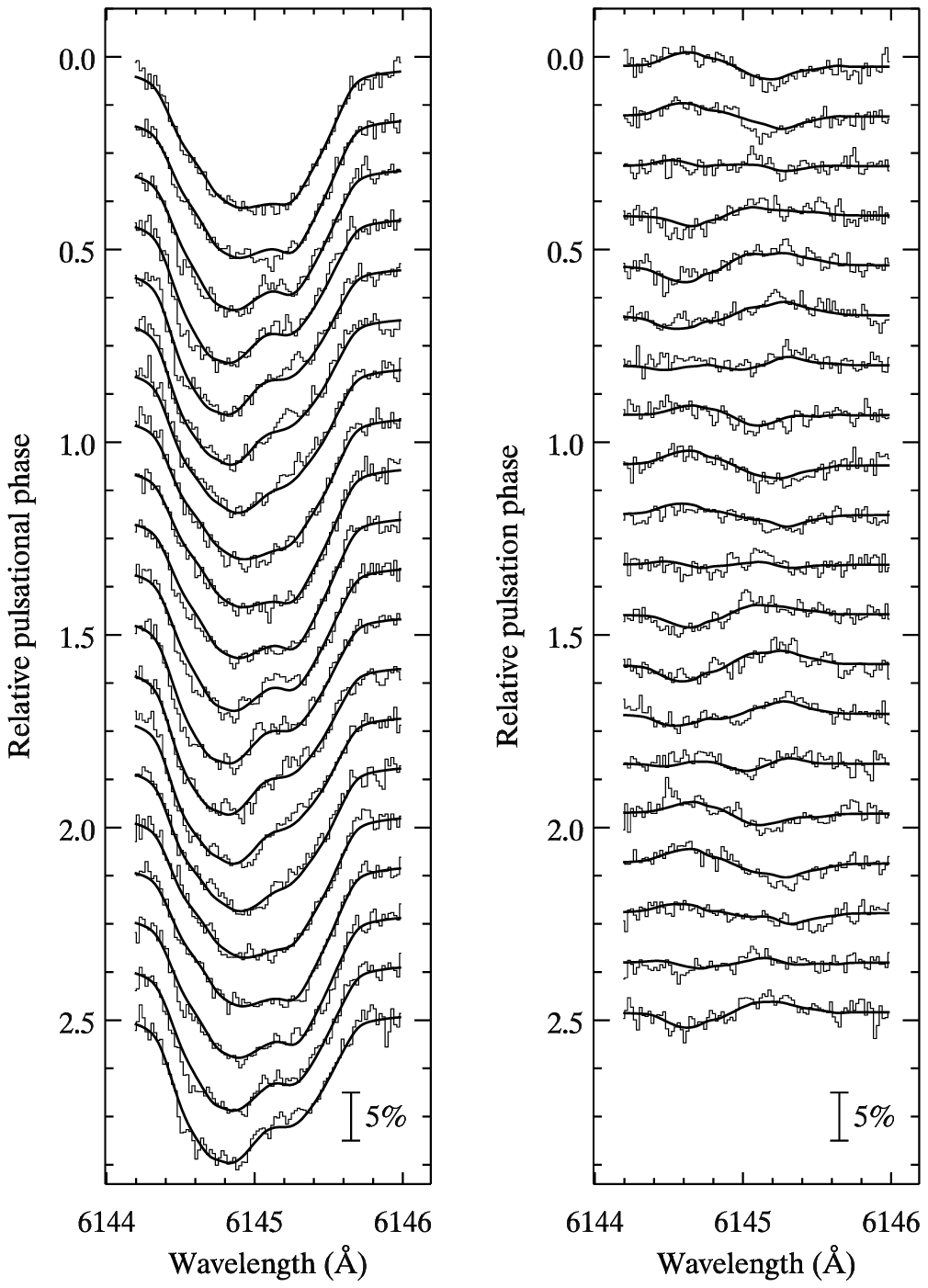}
\caption{
Observed variation of the \ion{Nd}{3} $\lambda$~6145~\AA\ line ({\it histogram}) is compared with the
best fit synthetic spectra ({\it solid line}). The left panel shows the 
observed and synthetic spectra
for 20 phases (out of 1860 used in the inversion) close to the
maximum of pulsational fluctuations (rotation phases $\varphi=0.966$-0.973 according to the ephemeris of
Kurtz et al. \citeyear{kwr97}). The right panel presents the difference between 
individual profiles and the average line shape. The spectra
for consecutive pulsation phases are shifted in the vertical direction. The bars
at the bottom of the panels show the vertical scale in units of the continuum
flux.\label{fig1}}
\end{figure}

The pulsation Doppler inversion is applied to time-resolved
observations of the \ion{Nd}{3} $\lambda$~6145~\AA\ line in the spectrum of the
well-known roAp star \hr\ (HD\,83368). A total of 1860 spectra of this object
were obtained over the period of 11 nights using the Coud\'e Echelle
Spectrograph fiber-linked to the Cassegrain focus of the 3.6-m telescope at the
European Southern Observatory. A preliminary discovery report of the
pulsational variations in individual spectral lines in \hr\ using these
observational data was published by \citet{kr01}. Our observations evenly
sample the 2\fd851976 rotation period of the star \citep{kwr97} and are
characterized by the resolving power of $\lambda/\Delta\lambda=123,000$ and the
signal-to-noise ratio of 110--160 pixel$^{-1}$. A 70$^{\rm s}$ exposure time
was chosen to ensure an appropriate sampling of the 11\fm67 
oscillation period. Fig.\,\ref{fig1} shows an example of the rapid line profile
variation of \hr\ and presents a comparison between the observed and
calculated spectra.

Reconstruction of the pulsation velocity field of \hr\ took into account an 
inhomogeneous surface distribution of neodymium which significantly distorts
the mean \ion{Nd}{3} line shapes. The maps
of chemical abundance and pulsational fluctuations were recovered
simultaneously in a self-consistent manner. The resulting Nd abundance map agrees
very well with the distribution derived by \citet{kdp04} using a different DI
code. 

Nonradial pulsations in roAp stars are characterized by very small
photometric amplitudes. The respective pulsational temperature variation 
does not exceed $\sim$\,10~K \citep{mk98}. Such temperature changes do not have
a noticeable effect on the stellar line profiles and can be safely neglected in
our modeling. 

The formation region of the \ion{Nd}{3} $\lambda$~6145~\AA\ line is located 
high in the atmospheres of roAp stars \citep{rpk02}. The horizontal motion due
to the high-overtone {\it p-}mode pulsation is expected to be either negligible
\citep{bd02} or considerably smaller \citep{sg04} than the radial motion at this
atmospheric height. Trial mapping of the full vector pulsation velocity
distribution demonstrated that, whereas including the horizontal pulsation
motion substantially increases the number of free parameters, it does not lead
to an appreciable improvement of the fit to observations. Taking this into
account we limited the pulsation DI reconstruction to the vertical velocity
component which is represented by the $V^c_r$ and $V^s_r$ velocity amplitudes.


The Doppler maps of the pulsation velocity field of \hr\ revealed with our technique
are illustrated in Fig.\,\ref{fig2}. The $V^c_r$ map shows a clear oblique dipolar
pattern with the local pulsation amplitude reaching up to 4.0~\kms. On the other hand,
the amplitude of the stuctures in the second map, $V^s_r$, does not exceed a few hundred
\ms. Such a relation of the two maps is typical of nearly axisymmetric pulsations
\citep[see Sect 5.5 of][]{k04}.


\section{Discussion}

Indirect imaging of the surface structure of nonradial oscillations has important
implications for stellar astrophysics. Results of the first Doppler inversion
of the pulsation velocity presented here can be regarded as an ultimate solution of the
long-standing problem of the pulsation geometry of roAp stars. Until now the
high-overtone {\it p-}mode oscillations in these unique objects were
investigated primarily with the high-speed photometric techniques. Analyses of
light variations of roAp stars \citep{k90} and, in particular, \hr\ 
\citep{kwr97}, demonstrated that the amplitude and phase of pulsations are
modulated with the stellar rotation. A coincidence of the times of maxima of pulsation
amplitude and magnetic field extrema was interpreted within the oblique
pulsator framework \citep{k82}. This phenomenological model postulated that
pulsations are represented by an axisymmetric dipolar ($\ell=1$) mode inclined
with respect to the stellar rotation axis and aligned with the axis of the
global magnetic field. Subsequent theoretical studies were carried out by
\citet{st93}, \citet{ts95} and, most recently, by \citet{sg04} assuming a
dominant role of the magnetic field. These calculations suggested that
important deviations of the mode geometry from the purely dipolar topology
might occur, although pulsation structure was still expected to remain almost
axisymmetric and aligned with the magnetic field. A totally different picture
has emerged from the theoretical investigation by \citet{bd02} who focused on
the essential role of the centrifugal force resulting from the stellar
rotation. According to their theory pulsations in roAp stars are described by a
superposition of all $\ell=1$ harmonic components and are neither axisymmetric, nor
necessarily aligned with the magnetic field. Limited information content of the
photometric observations did not allow astronomers to distinguish between these
competing approaches to the roAp pulsation structure. 

\begin{figure}
\epsscale{0.95}
\plotone{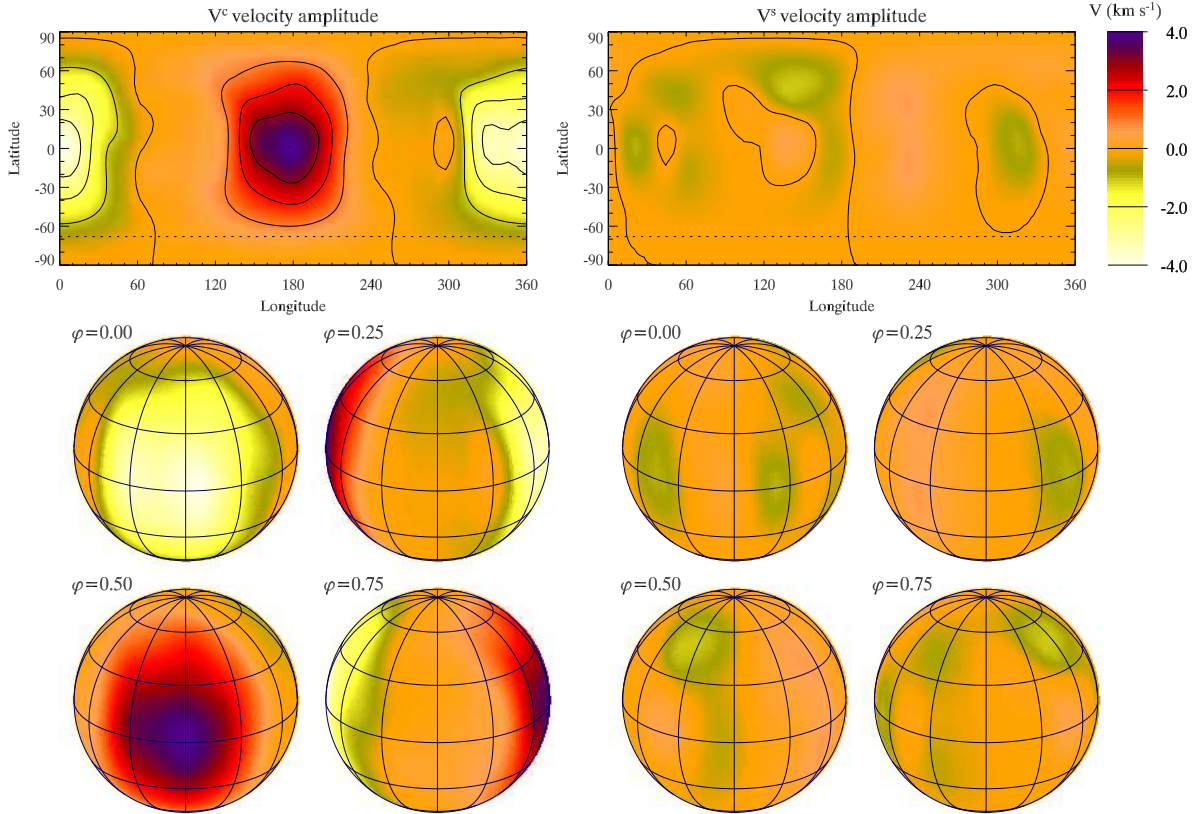}
\caption{
Results of the first Doppler imaging reconstruction of the stellar
pulsation velocity field. 
Rectangular panels show the $V^c_r$ ({\it left}) and $V^s_r$ ({\it right})
vertical velocity amplitude maps.
The contours of equal velocity are plotted over the greyscale images with a step of 1.0
\kms. The horizontal dotted line shows the lowest visible latitude
corresponding to the angle between the stellar axis of rotation and the
observer's line of sight (`inclination angle', $i=68\degr$) adopted in the
analysis of \hr. The plots below present spherical projection of the
respective pulsation velocity maps. The star is shown at 4 different aspect
angles corresponding to the rotation phases $\varphi=0.00$, 0.25, 0.50, and 0.75. 
The grid at the stellar surface is plotted with a 30\degr\ step in longitude and
latitude. 
\label{fig2}}
\end{figure}

The surface velocity mapping technique is not limited to the spherical
harmonic parameterization of pulsational fluctuations and can be used
to explore the physics of nonradial oscillations in rotating magnetic stars.
In particular, it is able to supply a crucial observational constraint required 
for testing the alternative theories of roAp pulsations. The pulsation map
(Fig.\,\ref{fig2}) reconstructed here for the prototype roAp star \hr\
displays an obvious correlation with the stellar magnetic field geometry.
\citet{kdp04} found that the dominant dipolar component of the global field
in \hr\ is inclined by about 90\degr\ relative to the stellar rotation axis
and is oriented at $\approx180\degr$ in longitude. At the same time, the
$V^c_r$ velocity map presented in Fig.\,\ref{fig2} indicates that pulsation
geometry of \hr\ is dominated by an axisymmetric structure. Furthermore, 
it is clear that the two maxima of the pulsation amplitude coincide with
the magnetic poles. This is the first independent verification of the
alignment of pulsations and magnetic field in oscillating magnetized
stars. 

Our pulsational mapping of \hr\ suggests a dominant role of the
magnetic perturbation of the {\it p-}modes and considerably less important
influence of the stellar rotation. Indeed, the pulsation Doppler image 
of the surface velocity field reveals no significant asymmetry which can be 
attributed to the oblique $\ell=1$, $|m|=1$ components.
Therefore, we conclude that pulsations in \hr\ cannot be described
by the theory brought forward by \citet{bd02}, who attempted to fully take into
account the rotationally induced mode perturbations but limited their
calculations to a relatively weak magnetic field.

\begin{figure}
\epsscale{0.7}
\plotone{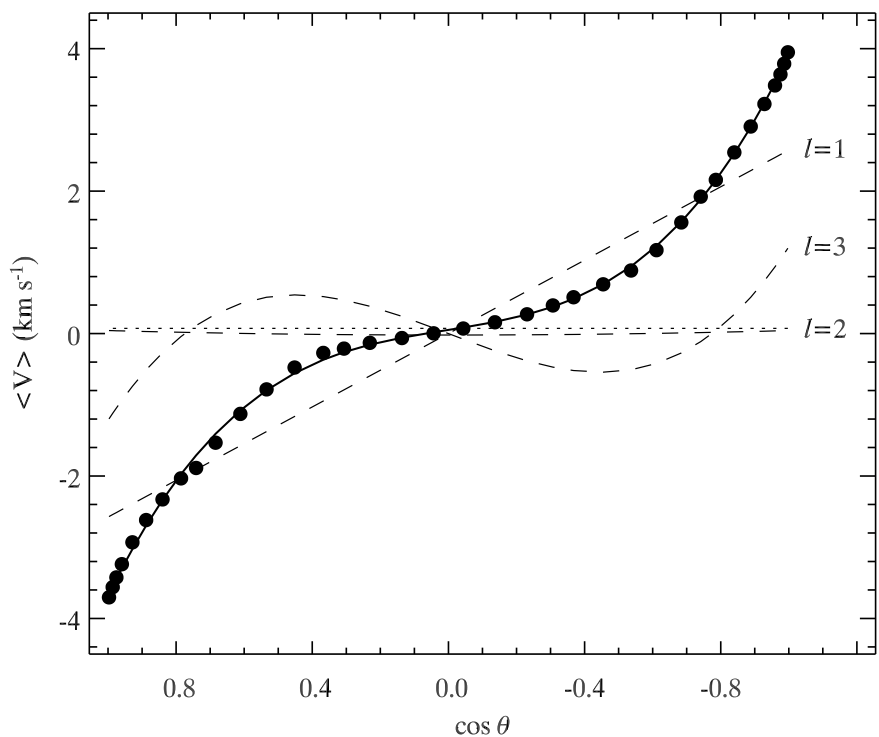}
\caption{
Analysis of the latitudinal dependence of the $V^c_r$ pulsation velocity
amplitude. The symbols show average
velocity amplitude as a function of the cosine of the angle $\theta$ counted from one
of the pulsation poles. The solid line illustrates the fit with a superposition
of axisymmetric $\ell=0$, 1, 2, and 3 spherical harmonic components. Contributions of
individual multipolar components are shown by the dashed lines for $\ell=1$--3 and by
the dotted line for $\ell=0$. \label{fig3}}
\end{figure}

Fig.\,\ref{fig3} illustrates an analysis of the dominant axisymmetric
pulsation structure reconstructed with our technique. This figure shows the
latitudinal dependence of the $V^c_r$ pulsation amplitude in the reference frame of
the stellar magnetic field. The inferred latitudinal trend is clearly
different from the one expected for a pure oblique $\ell=1$ pulsation. Axisymmetric
multipolar decomposition of the observed latitudinal dependence suggests
that oscillations in \hr\ can be roughly described with a superposition of
$\ell=1$ and $\ell=3$ harmonic components. The octupolar component has the
same sign as the dipolar contribution and an amplitude smaller by a factor
of two (A$_{\ell=3}$/A$_{\ell=1}$=$0.50\pm0.04$). In other words, we
discover that pulsations in \hr\ are strongly confined to the magnetic field
axis. This pulsation geometry is in remarkable agreement with the
predictions obtained in recent calculations assuming the dominant role of
the magnetic field \citep{sg04}. This concordance of the empirical model and theory
documents a significant breakthrough in our understanding of the physics of
{\it p-}mode oscillations in magnetic stars and possibly confirms an
important role of damping by the magnetic slow waves investigated by
\citet{sg04} in the context of excitation of the roAp-type pulsations.

In addition to revealing the true pulsation geometry of roAp stars, our
precise velocity map recovered for \hr\ opens numerous possibilities for
asteroseismological determination of various stellar parameters. For
instance, \citet{ts95} argued that the presence of a large quadrupolar
component in the magnetic field geometry is expected to induce $\ell=0$ and
$\ell=2$ axisymmetric structures in the pulsation velocity distribution.
However, Fig.\,\ref{fig3} convincingly demonstrates that the even multipolar
components are very weak in \hr.  Hence, the magnetic field of this star
must be basically dipolar -- a conclusion obtained independently from an
analysis of any classical magnetic field observables.

\acknowledgements
This work was supported by the Lise Meitner fellowship granted by the Austrian
Science Fund (FWF, project No. M757-N02). We acknowledge extensive use of 
the SIMBAD database, operated at CDS, Strasbourg, France.


\clearpage

\clearpage 

\clearpage 

\end{document}